\documentclass{emulateapj}

\slugcomment{To appear in ApJ Letters}

\shorttitle{Parsec-Scale Radio Emission from the LLAGN in He 2-10}
\shortauthors{Reines \& Deller}

\begin{document}

\title{Parsec-Scale Radio Emission from the Low-Luminosity Active Galactic \\ Nucleus in the Dwarf Starburst Galaxy Henize 2-10}

\author{Amy E. Reines\altaffilmark{1}}
\affil{National Radio Astronomy Observatory,
    Charlottesville, VA 22903}
\email{areines@nrao.edu}

\and

\author{Adam T. Deller}
\affil{The Netherlands Institute for Radio Astronomy (ASTRON), Dwingeloo, The Netherlands}
    
\altaffiltext{1}{Einstein Fellow}

\begin{abstract}

A candidate accreting massive black hole (BH) with $M_{\rm BH} \sim 10^6 M_\odot$ has recently been identified at the center of the dwarf starburst galaxy \object[He 2-10]{Henize 2-10} (He~2-10).  This discovery offers the first possibility of studying a growing BH in a nearby galaxy resembling those in the earlier universe, and opens up a new class of host galaxies to search for the smallest supermassive BHs.  Here we present very long baseline interferometry observations of He~2-10 taken with the Long Baseline Array (LBA) at 1.4~GHz with an angular resolution of $\sim 0\farcs1 \times 0\farcs03$.  A single compact radio source is detected at the precise location of the putative low-luminosity active galactic nucleus.  The physical size of the nuclear radio emission is $\lesssim$ 3~pc $\times$ 1~pc, an order of magnitude smaller than previous constraints from the Very Large Array (VLA), and the brightness temperature of $T_{\rm B} > 3 \times 10^5$~K confirms a non-thermal origin. These LBA observations indicate that the nuclear radio emission originates from a single object and exclude the possibility of multiple supernova remnants as the origin of the nuclear radio emission previously detected with the VLA at lower resolution.  A weaker, more extended, off-nuclear source is also detected with the LBA and a comparison with multi-wavelength ancillary data indicate that, unlike the nuclear source, the off-nuclear source is co-spatial with a super star cluster, lacks a detectable X-ray point-source counterpart, and is almost certainly due to a supernova remnant in the host star cluster.

\end{abstract}


\section{Introduction}\label{sec_intro}

Supermassive black holes (BHs) with $M_{\rm BH} \approx 10^6 - 10^9 M_\odot$ are thought to reside in the nuclei of essentially all massive galaxies with bulges \citep{Kormendy95,Magorrian98}, however the origin of these BHs is largely unknown.  Dwarf galaxies with low masses and relatively quiet merger histories are potential hosts of the least-massive BHs, and can therefore provide valuable constraints on the properties of the first primordial ``seed" BHs \citep[e.g.][]{Volonteri08,vanWassenhove10,Bellovary11}.  The BH mass distribution and occupation fraction in modern dwarfs can help distinguish between various seed formation mechanisms at high-redshift, such as stellar mergers in compact star clusters \citep[e.g.][]{Gurkan04,Devecchi09}, remnants from Population III stars \citep[e.g.][]{Madau01}, or the collapse of massive objects from proto-galactic gas \citep[e.g.][]{Bromm03,Begelman06,Lodato06}.  Active galactic nuclei (AGN) in dwarf galaxies undergoing significant star formation are particularly important for studying BH growth in physical conditions similar to those expected in the earlier universe. 

Observationally, however, few dwarf galaxies are known to host massive (as opposed to stellar-mass) BHs.  The most well-studied systems in the low-mass regime are the low-luminosity active galactic nuclei (LLAGN) in NGC~4395 \citep{Filippenko89,Filippenko03}, an essentially bulgeless late-type spiral galaxy with a nuclear star cluster, and Pox~52 \citep{Barth04,Thornton08}, a dwarf elliptical.  Estimates for the BH masses in these galaxies are $\sim {\rm few} \times 10^5 M_\odot$ \citep{Filippenko03,Barth04,Peterson05}.  Systematic searches for other low-mass BHs by \citet{Greene04, Greene07} have revealed $\sim 200$ BHs with virial masses of $M_{\rm BH} \sim 10^5 -10^6 M_\odot$ in a sample of broad-line AGN from the Sloan Digital Sky Survey.  Although most of the host galaxies are larger and brighter than typical dwarfs (the vast majority have extended disks), they are sub-$L^*$ galaxies without classical bulges \citep{Greene08, Jiang11}.  

\begin{figure}[!t]
\begin{center}$
\begin{array}{c}
{\includegraphics[width=3.2in]{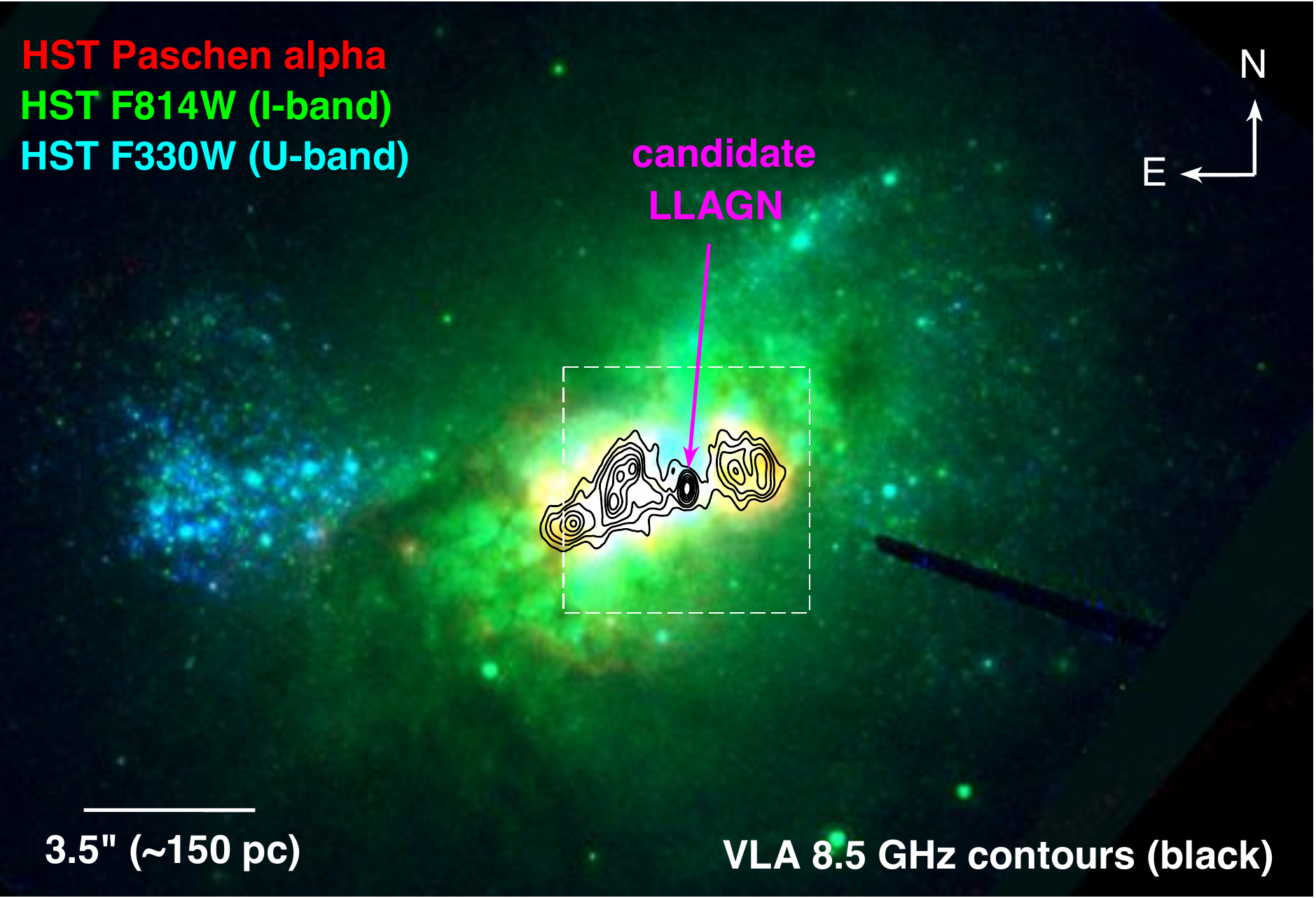}} 
\end{array}$
\end{center}
\caption{Three-color {\it HST} image of He~2-10 with 8.5 GHz VLA contours from \citet{Reines11}.  The dashed box indicates the field of view of the LBA image shown in Figure \ref{fig:vlbi}. 
\label{fig:he210}}
\end{figure}

Recently, \citet{Reines11} have discovered a candidate accreting massive BH at the center of \object[He 2-10]{Henize 2-10} (He~2-10), a nearby (9 Mpc), gas-rich, dwarf galaxy in the midst of an intense burst of star formation \citep[e.g.][]{Allen76,Johansson87,Kobulnicky95,Johnson03}. 
He~2-10 has a compact ($\sim 1$~kpc) irregular morphology (Figure \ref{fig:he210}) with no discernible bulge, and hosts a large population of young, massive and dense ``super star clusters" that are thought to be progenitors of globular clusters \citep{Conti94,Johnson00,Chandar03}.  In fact, the study of infant massive star clusters \citep{Reines08a, Reines08b, Reines10} in He~2-10 is what ultimately led to the serendipitous discovery of the putative BH \citep{Reines11}.

The LLAGN is detected as a central unresolved radio source with the Very Large Array \citep[VLA\footnote{The National Radio Astronomy Observatory is a facility of the National Science Foundation operated under cooperative agreement by Associated Universities, Inc.};][]{Kobulnicky99,Johnson03,Reines11} that is spatially coincident with a hard X-ray point source observed with the {\it Chandra X-ray Observatory} \citep{Ott05,Kobulnicky10,Reines11} and clearly separated from the prominent H{\footnotesize II} regions in the galaxy.  No star cluster is seen at the location of the nuclear radio source even in deep near-infrared {\it Hubble Space Telescope (HST)} observations \citep{Reines11}, and the position of the radio source is consistent with the dynamical center of the galaxy.  Moreover, the nuclear radio source lies at the center of a $\sim 250$~pc-long ionized gas structure with a velocity gradient that is consistent with an outflow or rotating disk of ionized material.  The relative strength (i.e.\ the ratio) of the radio to hard X-ray luminosities of the compact central source is consistent with known LLAGN \citep{Ho08}, but differs by orders of magnitude from active X-ray binaries and supernova remnants (SNRs).  A young supernova origin is ruled out since the VLA flux density at 8.5~GHz has not measurably declined in more than a decade \citep[from 1994 to 2004;][]{Johnson03,Reines11} and Very Long Baseline Array (VLBA) observations at 5 GHz do not detect any sources on scales of $\sim 0.1$ pc \citep[a typical size for a young compact radio supernova;][]{Ulvestad07}.  The VLBA non-detection of the nuclear VLA source, however, does not rule out a more extended LLAGN jet.  Using an empirical relationship between compact radio emission, compact hard X-ray emission and BH mass \citep[the ``fundamental plane of black hole activity,''][]{Merloni03}, Reines et al. (2011) estimate a mass of log$(M_{\rm BH}/{M}_\odot) = 6.3 \pm 1.1$ for the BH in He~2-10.

As part of an effort to verify the nuclear source in He~2-10 is indeed a LLAGN, we are undertaking follow-up observations of the nuclear radio emission using very long baseline interferometry (VLBI).  In this Letter, we present VLBI observations of He~2-10 taken with the Long Baseline Array (LBA) at 1.4 GHz that have a beam solid angle $\sim 38 \times$ smaller than the VLA observations of \citet{Reines11} and $\sim 128 \times$ larger than the VLBA observations of \citet{Ulvestad07} that apparently resolved out the nuclear source.  Our LBA observations reveal a parsec-scale radio source coincident with the candidate LLAGN, supporting the case for a massive BH in He~2-10.

\section{Observations and Data Reduction}

\begin{figure}[!t]
\begin{center}$
\begin{array}{cc}
{\includegraphics[width=3.2in]{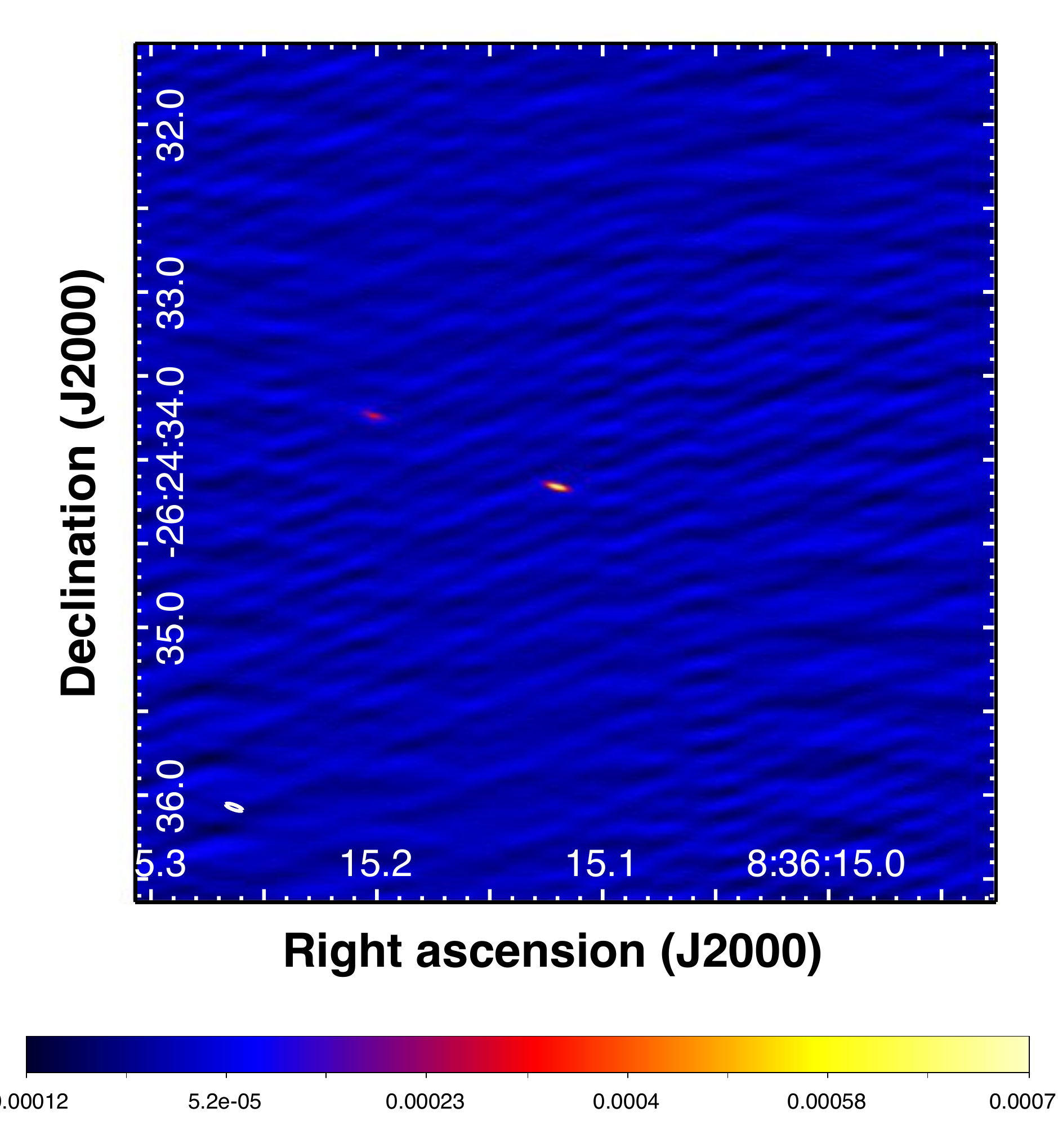}} \\
{\includegraphics[width=3.2in]{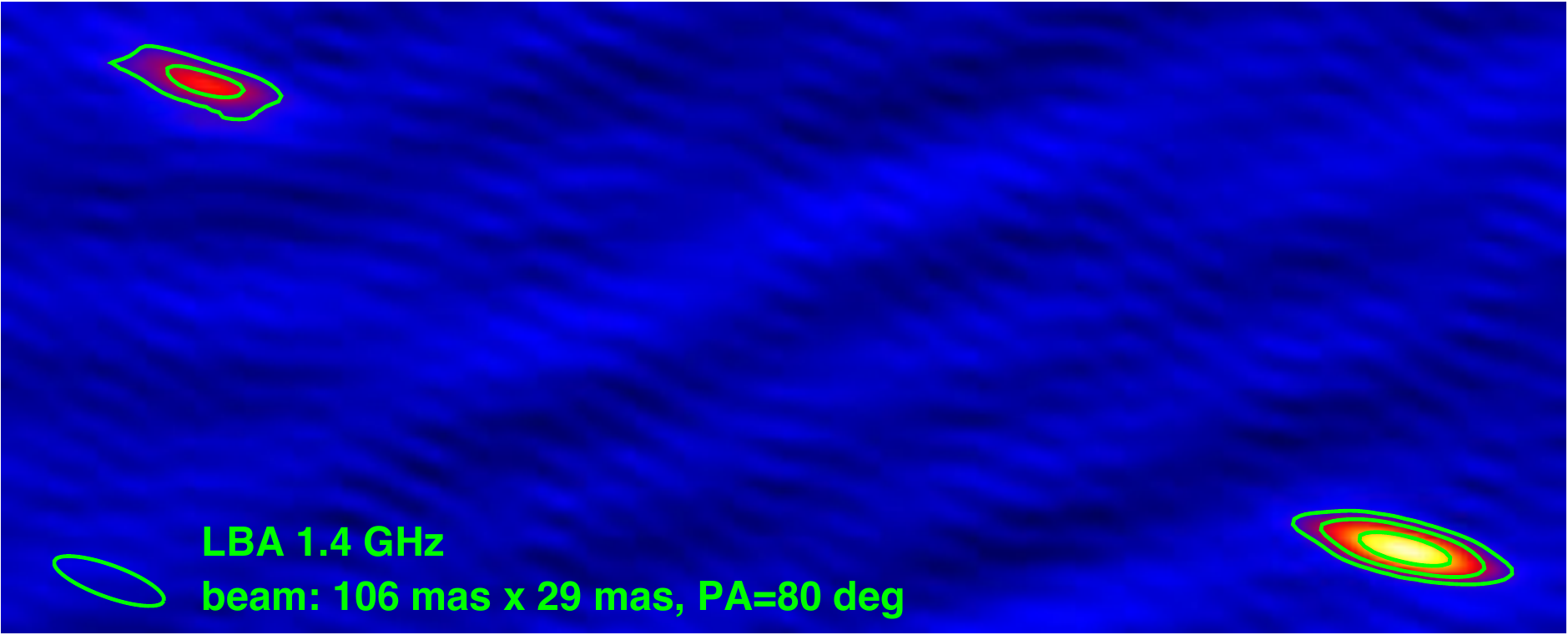}}
\end{array}$
\end{center}
\caption{{\it Top:} Final cleaned LBA image of He~2-10.  The brighter central source is the candidate LLAGN.  The beam (106~mas $\times$ 29~mas) is shown in the lower left corner.  
{\it Bottom:} A close-up view of the two detected sources.  Contour levels are $4, 8~{\rm and}~16 \times$ the rms noise $(1 \sigma = 32 \mu$Jy/beam).  The colorbar has units of Jy/beam. 
\label{fig:vlbi}}
\end{figure}

\begin{deluxetable*}{lrr}
\tabletypesize{\tiny}
\tablecaption{Single Gaussian Model Fits}
\tablewidth{0pt}
\tablehead{
\colhead{Parameter} & \colhead{Nuclear source} & \colhead{Off--nuclear source}}
\startdata
Right Ascension (J2000)		&  	08$^{\mathrm h}$36$^{\mathrm m}$15.1200$^{\mathrm s}$ $\pm$ 0.0003
						&	08$^{\mathrm h}$36$^{\mathrm m}$15.2016$^{\mathrm s}$ $\pm$ 0.0004 	\\
Declination (J2000) 			&  	$-$26\degr$24'34.157''$ $\pm$ 0.004	
						&	$-$26\degr$24'33.732''$ $\pm$ 0.005 \\
Integrated flux density (mJy)	&	0.98 $\pm$ 0.06 $\pm$ 0.20	&	0.59 $\pm$ 0.08 $\pm$ 0.12 \\
Peak flux density (mJy/beam)	&  	0.73 $\pm$ 0.03 $\pm$ 0.15 	&	0.29 $\pm$ 0.03 $\pm$ 0.06 \\
Fitted Major axis (milliarcseconds)		&	124 $\pm$ 5	&137 $\pm$ 13\\
Fitted Minor axis (milliarcseconds)	&	33 $\pm$ 1	& 	44 $\pm$ 4 \\
Fitted Position angle (degrees)		& 	77$^{+1}_{-1}$		& 71$^{+3}_{-3}$ \\
Deconvolved Major axis (milliarcseconds)	&	65 $\pm$ 10		& 	90 $\pm$20 \\
Deconvolved Minor axis (milliarcseconds)	&	10 $\pm$ 10		& 	20 $\pm$20 \\
Deconvolved Position angle (degrees)		& 	68$^{+7}_{-4}$		&	60$^{+15}_{-9}$ \\
\enddata
\tablecomments{Flux densities are quoted as fitted value $\pm$ fit error $\pm$ absolute flux density scale error (estimated at 20\%).  At the adopted distance to He~2-10 (9 Mpc), 10 mas $\approx$ 0.44 pc.}
\label{tab:sources}
\end{deluxetable*}

We observed He~2-10 on 2011 July 22 with the LBA in eVLBI mode \citep{Phillips07}.  Continuum observations at 1.4 GHz (21 cm) were made using the four eVLBI capable antennas at this frequency -- the Australia Telescope National Facility (ATNF) observatories at Parkes (64-m single dish), the Australia Telescope Compact Array (ATCA; five 22-m antennas phased), and Mopra (22-m single dish), as well as the Hobart antenna (26-m single dish) operated by the University of Tasmania.  The observation duration was 8 hours, but the target source was only visible to Parkes for 5 hours, due to the relatively restrictive Parkes elevation limit.  Two bands of width 64 MHz were sampled with two bit precision in the frequency ranges 1.368 -- 1.432 GHz and 1.432 -- 1.496 GHz at the three ATNF stations, while only the lower frequency band was sampled at Hobart due to the more limited network connectivity to this station.  The data (with an aggregate rate of 3.5 gigabits per second) were correlated on a computer cluster at the ATCA in real time using the DiFX software correlator \citep{Deller07}.
 
The observations utilized a phase reference cadence of 5-minute loops with 3.5 minutes on-source and 1.5 minutes on a phase calibrator separated by 2.4$\degr$ (J0846-2607).  Additionally, a fringe finder (0537-441 or 0834-291) was observed for 5 minutes before and after every $\sim$2-hour block on He~2-10 and the phase calibrator.  After editing, a total of $\sim$3.5 hours of usable integration time was obtained on He~2-10, of which the first 50 minutes lacked the Parkes antenna.  The predicted sensitivity for the target field using natural weighting was 28 $\mu$Jy/beam.  However, data from the phase reference source showed that the phased ATCA station suffered from severely degraded sensitivity during the first half of the experiment (reduced by a factor of $\sim$3). Instrumental failure or poor phasing of the array are potential explanations, but the actual cause could not be determined based on system logs and monitor data.  As the ATCA is the second most sensitive station in the array, this corresponds to a significant (10\%) reduction in the predicted final image sensitivity to 31 $\mu$Jy/beam.

Data reduction was performed using the AIPS\footnote{http://www.aips.nrao.edu/index.shtml} 
software package, and imaging was performed using Difmap \citep{Shepherd97}. The data were first amplitude calibrated using the logged system temperature data, delay calibrated using the phase reference source, and then phase and amplitude calibration were performed using a model of J0846-2607 derived from these observations.  The absolute amplitude calibration accuracy of the visibility data is expected to be at the 20\% level.  The target field data were split, averaged in frequency and exported from AIPS in uvfits format for imaging in Difmap.

In Difmap, a $1024 \times 1024$ pixel image with a pixel size of 5 milliarcseconds (mas) was formed using natural weighting.  Two sources were detected and cleaned.  After cleaning, a phase self--calibration was applied with a solution interval of 15 minutes, the shortest time at which reliable solutions could be obtained. This self--calibration step increased the recovered flux density of the two sources by $\sim$40\%.  Figure~\ref{fig:vlbi} shows the final cleaned image.  The 1$\sigma$ image sensitivity is 32~$\mu$Jy/beam, consistent with expectations, and the angular resolution is 106~mas $\times$ 29~mas, at a position angle of 80\degr.  At the adopted distance of He~2-10 (9 Mpc), this angular resolution corresponds to a linear size of  $\sim$~4.6~pc $\times$ 1.3~pc.

To facilitate the comparison of multi-wavelength images of He~2-10, we register our LBA image and the data presented in \citet{Reines11} to the same astrometric reference frame.  We make the reasonable assumption that the two LBA sources coincide with two compact, non-thermal VLA sources to derive the astrometric shift.  The sources in \citet{Reines11} have positions referenced to the Two Micron All Sky Survey catalog \citep{Skrutskie06} and are offset by $\sim 0\farcs 04$ west and $\sim 0\farcs 09$ north relative to the LBA positions.

\section{Results}\label{results}

We fit each of the two detected sources with a single Gaussian model using the task JMFIT in AIPS, the results of which are shown in Table~\ref{tab:sources}.  

\subsection{The Nuclear Source: LLAGN}\label{sec:nucleus}

The candidate LLAGN identified by \citet{Reines11} is clearly detected in our LBA observations at 1.4 GHz and is compact on parsec scales.  Although the source appears marginally resolved, the significance of this measurement is low.  In addition, residual phase and amplitude calibration errors will certainly remain after phase referencing and self--calibration, since the atmospheric phase will vary on timescales shorter than the solution interval of 15 minutes, and the signal--to--noise of the calibration solutions is relatively low.  These residual errors will have the effect of broadening the point--source response, and so we consider the deconvolved source size (plus errors) to be an upper limit on the actual source size.  This upper limit of 75 mas $\times$ 20 mas corresponds to a linear size $\lesssim$ 3 pc $\times$ 1 pc.  Compared to the VLA observations of \citet{Reines11}, which have a linear resolution of $\sim 24~{\rm pc} \times 9~{\rm pc}$, this yields an order-of-magnitude reduction in the maximum source size.

\begin{figure*}[!t]
\begin{center}$
\begin{array}{cc}
{\includegraphics[width=3.35in]{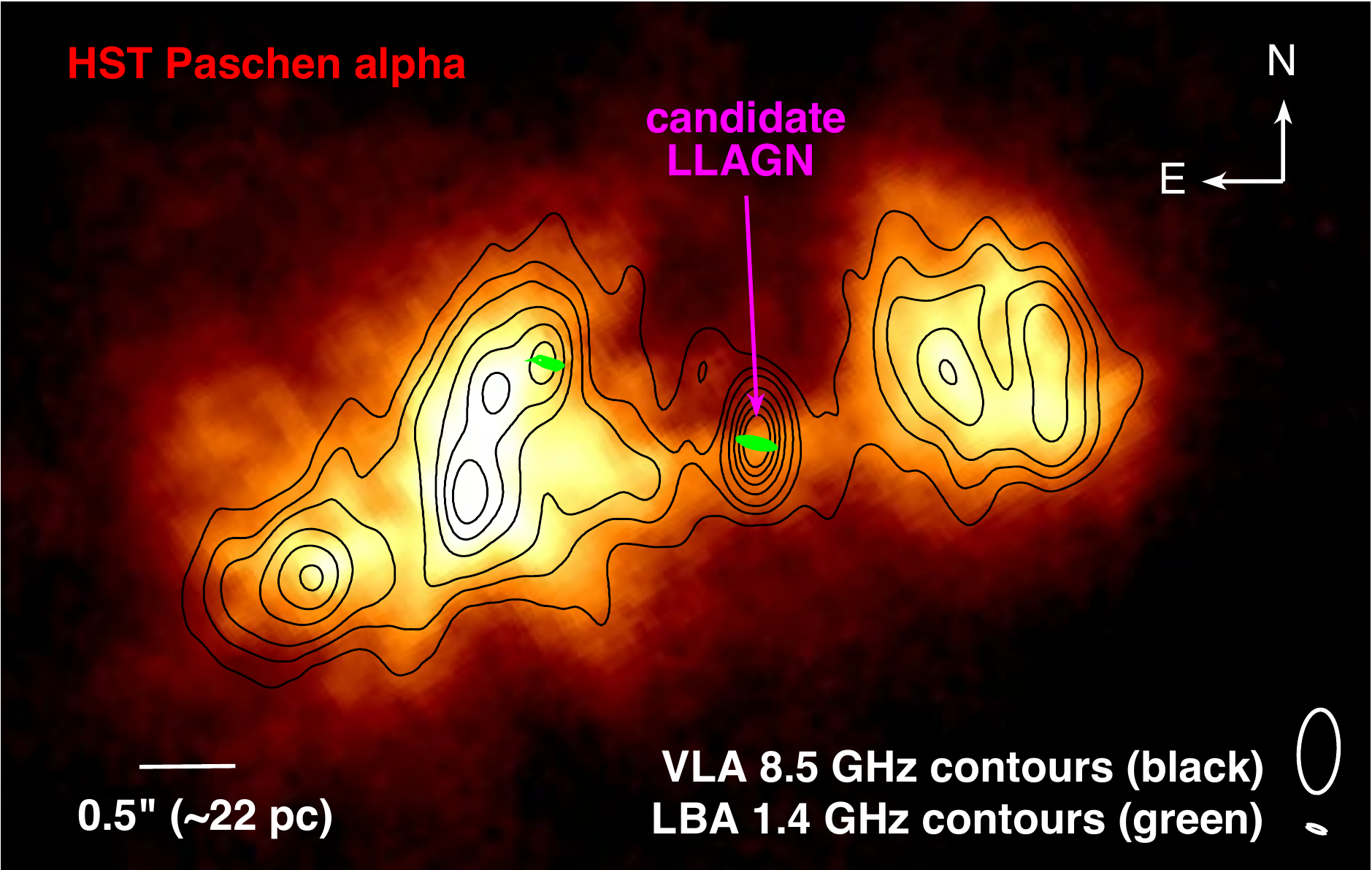}} &
{\includegraphics[width=3.35in]{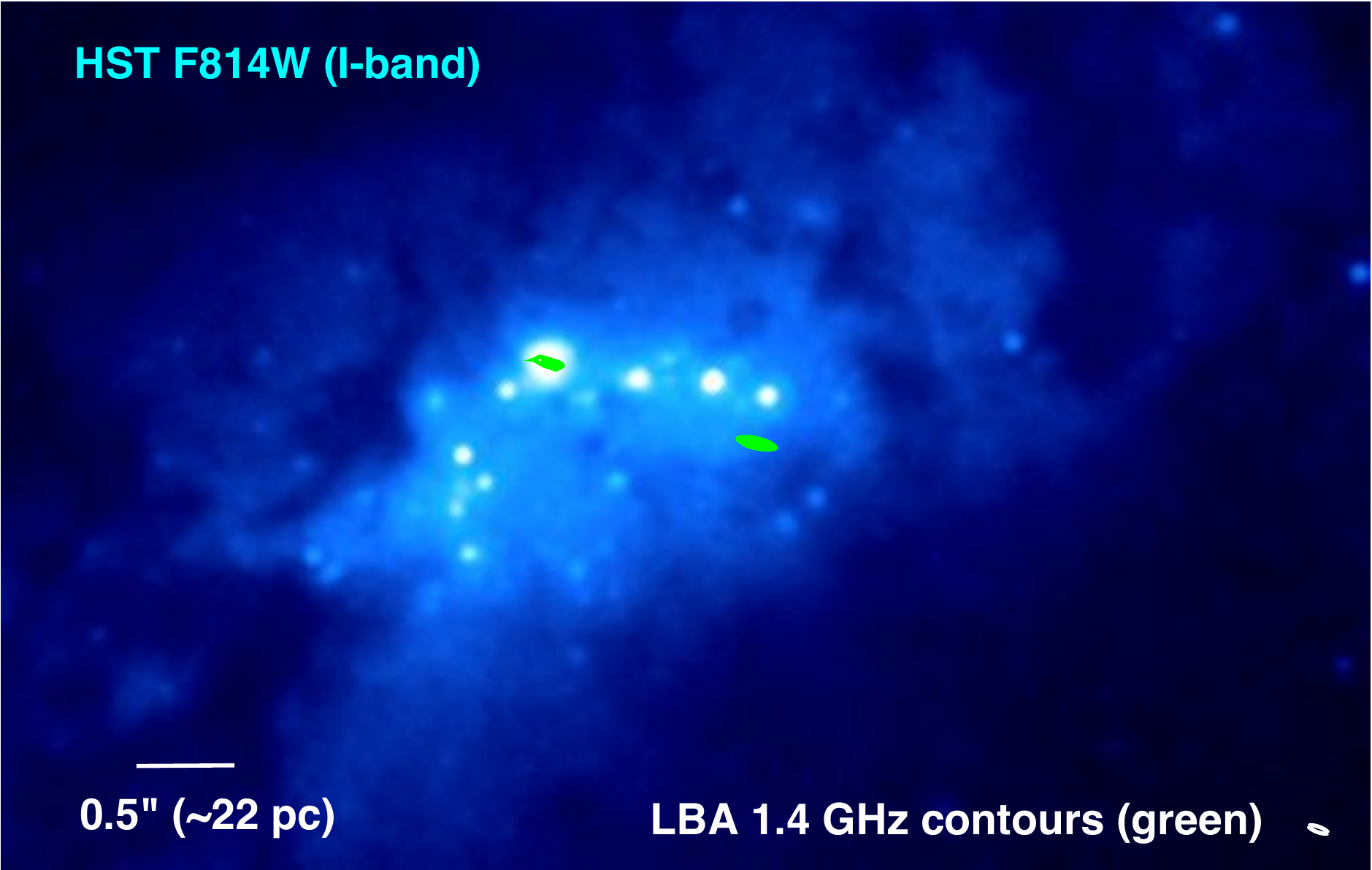}}
\end{array}$
\end{center}
\caption{{\it Left:} {\it HST} Pa$\alpha$ image from \citet{Reines11} of the central region of He~2-10 showing ionized gas emission.  Radio contours are in green (LBA) and black (VLA) and the beam sizes are shown in the lower right corner.  The candidate LLAGN is at the center of the $\sim 250$ pc-long ionized gas structure.   
{\it Right:} {\it HST} F814W ($I$-band) image of the same region showing young super star clusters.  The off-nuclear LBA source is spatially coincident with the most luminous stellar cluster and is almost certainly related to a SNR.  
\label{fig:multiwave}}
\end{figure*}

The lower-resolution VLA observations allowed for the possibility of multiple discrete sources contributing to the flux density in the nuclear region, such as a collection of SNRs \citep{Ulvestad07,Reines11} with each remnant having a typical size of $\sim$ few pc \citep[e.g.\ M82;][]{Fenech10}.  However, we only detect one parsec-scale source at the center of this region (at the $\approx 31\sigma$ level), despite the LBA observations having a beam solid angle $\sim 38 \times$ smaller than the VLA observations.  This indicates the nuclear radio emission comes from a single object and argues against multiple SNRs within the unresolved nuclear VLA source.  It is also very unlikely that multiple SNRs reside within the nuclear LBA parsec-size source, given that VLBI observations of M82 reveal dozens of well-resolved SNRs that are clearly separated from one another and spread over the central kpc of the galaxy \citep[e.g.][]{Fenech10}. 

Using the fitted peak flux density of 0.73 mJy/beam and the synthesized beam size of 106 mas $\times$ 29 mas, we calculate a lower limit to the brightness temperature of $T_B > 1.4 \times 10^5$~K.  However, as described above, the emitting region is likely significantly smaller than the beam size.  Therefore, following \citet{Blundell96}, the lower limit on the brightness temperature can be increased to $T_B >  2.9 \times 10^5$~K using the upper limit on the source size of 75 mas $\times$ 20 mas.  This high brightness temperature rules out free-free emission from thermal electrons with a characteristic temperature of $\sim 10^4$~K (i.e.\ an H{\footnotesize II} region), and confirms a non-thermal origin for the nuclear radio source as indicated by the spectral index derived from VLA observations at higher frequencies and lower angular resolution \citep[][]{Johnson03,Reines11}.  

The integrated flux density of the nuclear LBA source is $0.98 \pm 0.21$ mJy.  At a distance of 9 Mpc, this corresponds to a spectral luminosity of $L_{1.4{\rm GHz}} \sim 10^{19}$ W Hz$^{-1}$.  For comparison, this is approximately an order of magnitude more luminous than the LLAGN in NGC 4395 at similar spatial scales (i.e.\ intermediate of VLA and VLBA observations of NGC 4395; Figure 2 in \citealt{Wrobel01}).  

The measured flux density is roughly consistent with predictions based on higher frequency VLA observations \citep{Reines11}.  We have re-derived the VLA flux densities using the AIPS task JMFIT to be consistent with the method used here, fitting a point source plus a background (level + slope) to the region immediately surrounding the nuclear source. We obtain flux densities of $0.86 \pm 0.02$ and $0.64 \pm 0.01$ mJy at 4.9 and 8.5 GHz, respectively.  These values are somewhat lower than those quoted in \citet{Reines11}, which were derived using larger apertures and different background regions (see the Supplementary Information).  We note, however, that this does not affect the results of that paper in any appreciable way.  Assuming an additional 15\% uncertainty for the absolute flux calibration of each VLA image, the spectral index $\alpha = -0.54 \pm 0.39$ ($S_\nu \propto \nu^\alpha$).  The predicted 1.4 GHz flux density is then $1.7^{+1.0}_{-0.7}$ mJy.  Thus, the integrated LBA 1.4 GHz flux density of $0.98 \pm 0.21$ mJy is (marginally) consistent with expectations if the spectral index remains constant between 1.4 and 8.5 GHz.  However, our data do not exclude the possibility that additional flux is present at 1.4 GHz on spatial scales intermediate between the LBA and VLA beams, similar to what is observed in steep-spectrum Seyfert nuclei in which the radio emission is thought to be dominated by low-surface brightness features such as jets \citep{Orienti10}.  It is also possible that the radio spectral energy distribution flattens at lower frequencies and exhibits a spectral turnover, a common characteristic of AGN cores \citep[e.g.][]{Pearson92, Odea98}.

\subsection{The Off-Nuclear Source: Supernova Remnant}

A second source is detected in our LBA observations that is located $\sim$1\arcsec\ to the north-east of the nucleus (Figure \ref{fig:vlbi}).  This source is fainter and more extended than the nuclear source, with $T_B \approx 4.0 \times 10^4$~K.  The brightness temperature of the nuclear source is therefore $\sim 7\times$ larger than the off-nuclear source.  Additionally, in contrast to the nuclear source, the off-nuclear source lacks a detectable X-ray point-source counterpart \citep{Ott05,Kobulnicky10,Reines11} and is spatially coincident with a luminous super star cluster (Figure \ref{fig:multiwave}), strongly suggesting the radio emission is from a SNR in the host cluster.  The most massive O stars in a super star cluster are expected to begin exploding as supernovae at an age of $\sim 3.5$ Myr \citep{Leitherer99} and the host star cluster is the most luminous in the galaxy with a mass of $\sim 2 \times 10^5 M_\odot$ and an age of $\sim 5$ Myr \citep{Chandar03}.   For comparison, the 1.4 GHz radio luminosity of the off-nuclear LBA source in He~2-10 is about twice that of the Galactic SNR Cas A, assuming the flux density and spectral index measured by \citet{Baars77} and the distance derived by \citet{Reed95}. 

\newpage

\section{Conclusions and Discussion}

We observed He~2-10 with the LBA at 1.4 GHz with an angular resolution of $\sim 0\farcs1 \times 0\farcs03$, corresponding to a linear size of  $\sim$~4.6~pc $\times$ 1.3~pc.  Our findings are summarized below:

\begin{enumerate}

\item We clearly detect compact radio emission from the candidate LLAGN identified by \citet{Reines11}, supporting the case for an accreting massive BH in this dwarf starburst galaxy. The physical extent of the nuclear radio emission is $\lesssim$3~pc $\times$ 1~pc, an order of magnitude smaller than previous constraints from the VLA.

\item Our observations indicate the nuclear radio emission originates from a single object.  The possibility of multiple SNRs as the origin of the nuclear radio emission previously detected by the VLA on larger scales is disfavored by the detection of only one bright compact LBA source in the nuclear region.

\item The high brightness temperature of the nuclear source, $T_B > 3 \times 10^5$~K, rules out an H{\footnotesize II} region and confirms a non-thermal origin.  

\item The integrated flux density of the nuclear LBA source at 1.4 GHz is $\sim 1$~mJy, corresponding to a spectral luminosity of $L_{1.4{\rm GHz}} \sim 10^{19}$ W Hz$^{-1}$.  This is an order of magnitude larger than the LLAGN in NGC 4395 at similar spatial scales.  

\item We detect a weaker and more extended off-nuclear source that, unlike the nuclear source, lacks a detectable X-ray point-source counterpart, is co-spatial with a super star cluster, and is almost certainly due to the emission from a SNR in the host star cluster.  

\end{enumerate}

While multi-wavelength observations provide a compelling case for the presence of a LLAGN in He~2-10  \citep[see \S\ref{sec_intro} and][]{Reines11}, even radio observations {\it alone} are placing stringent constraints on the nuclear source.  An H{\footnotesize II} region is ruled out by the high brightness temperature derived from the LBA observations presented here and the spectral index derived from previous VLA observations \citep{Johnson03,Reines11}.  A young compact supernova is ruled out by the VLBA non-detection at 5 GHz \citep{Ulvestad07} and the lack of flux evolution in 8.5 GHz observations from the VLA.  Multiple SNRs are ruled out by our LBA observations that detect only one parsec-scale nuclear source.  This leaves only two remaining possibilities for the nuclear radio emission: a LLAGN or a single SNR at the center of the galaxy, far from any detectable signs of star formation.  The luminosity of the nuclear source would be high for a SNR -- it is more powerful than 31 out of 36 known radio SNRs in M82 -- and it would likely be just decades old \citep{Fenech10}.  Future observations with the LBA (at higher frequency) and the VLBA (at low frequency) can be employed to further characterize the nuclear radio source in He~2-10, providing a high-resolution radio spectral energy distribution and detailed morphological information necessary for a secure identification.

\begin{center}
\end{center}

\acknowledgments

We thank Jim Condon, Kelsey Johnson and David Nidever for providing comments on the manuscript and useful discussions.  We are also grateful to Kelsey Johnson for obtaining the VLA observations used in this work, to Chris Phillips for providing assistance with the correlation of the LBA data set, and to the referee for a helpful report.  The LBA is part of the Australia Telescope which is funded by the Commonwealth of Australia for operation as a National Facility managed by CSIRO.  Support for A.E.R. was provided by NASA through the Einstein Fellowship Program, grant PF1-120086.

{\it Facilities:} \facility{LBA}, \facility{VLA}


\begin{thebibliography}{}

\bibitem[Allen et al.(1976)]{Allen76} Allen, D.~A., Wright, 
A.~E., \& Goss, W.~M.\ 1976, \mnras, 177, 91 

\bibitem[Baars et al.(1977)]{Baars77} Baars, J.~W.~M.,
Genzel, R., Pauliny-Toth, I.~I.~K., \& Witzel, A.\ 1977, \aap, 61, 99

\bibitem[Barth et al.(2004)]{Barth04} Barth, A.~J., Ho, L.~C., 
Rutledge, R.~E., \& Sargent, W.~L.~W.\ 2004, \apj, 607, 90 

\bibitem[Begelman et al.(2006)]{Begelman06} Begelman, M.~C., 
Volonteri, M., \& Rees, M.~J.\ 2006, \mnras, 370, 289 

\bibitem[Bellovary et al.(2011)]{Bellovary11} Bellovary, J., 
Volonteri, M., Governato, F., et al.\ 2011, \apj, 742, 13 

\bibitem[Blundell et al.(1996)]{Blundell96} Blundell, K.~M., 
Beasley, A.~J., Lacy, M., \& Garrington, S.~T.\ 1996, \apjl, 468, L91 

\bibitem[Bromm \& Loeb(2003)]{Bromm03} Bromm, V., \& Loeb, A.\ 2003, \apj, 596, 34 

\bibitem[Chandar et al.(2003)]{Chandar03} Chandar, R., Leitherer, 
C., Tremonti, C., \& Calzetti, D.\ 2003, \apj, 586, 939 

\bibitem[Conti \& Vacca(1994)]{Conti94} Conti, P.~S., \& Vacca, W.~D.\ 1994, \apjl, 423, L97 

\bibitem[Deller et al.(2007)]{Deller07} Deller, A.~T., Tingay, 
S.~J., Bailes, M., \& West, C.\ 2007, \pasp, 119, 318 

\bibitem[Devecchi \& Volonteri(2009)]{Devecchi09} Devecchi, B., \& Volonteri, M.\ 2009, \apj, 694, 302 

\bibitem[Fenech et al.(2010)]{Fenech10} Fenech, D., Beswick, R., 
Muxlow, T.~W.~B., Pedlar, A., \& Argo, M.~K.\ 2010, \mnras, 408, 607 

\bibitem[Filippenko \& Ho(2003)]{Filippenko03} Filippenko, A.~V.,
\& Ho, L.~C.\ 2003, \apj, 588, L13

\bibitem[Filippenko \& Sargent(1989)]{Filippenko89} Filippenko, A.~V., \& Sargent, W.~L.~W.\ 1989, \apjl, 342, L11 

\bibitem[Greene \& Ho(2004)]{Greene04} Greene, J.~E., \& Ho, L.~C.\ 2004, \apj, 610, 722 

\bibitem[Greene \& Ho(2007)]{Greene07} Greene, J.~E., \& Ho, L.~C.\ 2007, \apj, 670, 92 

\bibitem[Greene et al.(2008)]{Greene08} Greene, J.~E., Ho, 
L.~C., \& Barth, A.~J.\ 2008, \apj, 688, 159 

\bibitem[G{\"u}rkan et al.(2004)]{Gurkan04} G{\"u}rkan, M.~A., 
Freitag, M., \& Rasio, F.~A.\ 2004, \apj, 604, 632 

\bibitem[Ho(2008)]{Ho08} Ho, L.~C.\ 2008, \araa, 46, 475 

\bibitem[Jiang et al.(2011)]{Jiang11} Jiang, Y.-F., Greene, 
J.~E., Ho, L.~C., Xiao, T., \& Barth, A.~J.\ 2011, \apj, 742, 68 

\bibitem[Johansson(1987)]{Johansson87} Johansson, I.\ 1987, \aap, 182, 179 

\bibitem[Johnson \& Kobulnicky(2003)]{Johnson03} Johnson, K.~E., \&
Kobulnicky, H.~A.\ 2003, \apj, 597, 923 

\bibitem[Johnson et al.(2000)]{Johnson00} Johnson, K.~E., 
Leitherer, C., Vacca, W.~D., \& Conti, P.~S.\ 2000, \aj, 120, 1273 

\bibitem[Kobulnicky et al.(1995)]{Kobulnicky95} Kobulnicky,
H.~A., Dickey, J.~M., Sargent, A.~I., Hogg, D.~E., \& Conti, P.~S.\
1995, \aj, 110, 116

\bibitem[Kobulnicky \& Johnson(1999)]{Kobulnicky99} Kobulnicky, H. A. \& Johnson, K. E. 1999,
\apj, 527, 154

\bibitem[Kobulnicky \& Martin(2010)]{Kobulnicky10} Kobulnicky, H.~A., \& Martin, C.~L.\ 2010, \apj, 718, 724

\bibitem[Kormendy  \& Richstone(1995)]{Kormendy95}
Kormendy, J., \& Richstone, D.\ 1995, \araa, 33, 581 

\bibitem[Leitherer et al.(1999)]{Leitherer99} Leitherer, C., et 
al.\ 1999, \apjs, 123, 3 

\bibitem[Lodato \& Natarajan(2006)]{Lodato06} Lodato, G., \& Natarajan, P.\ 2006, \mnras, 371, 1813 

\bibitem[Madau \& Rees(2001)]{Madau01} Madau, P., \& Rees, M.~J.\ 2001, \apjl, 551, L27 

\bibitem[Magorrian et al.(1998)]{Magorrian98} Magorrian, J., et 
al.\ 1998, \aj, 115, 2285 

\bibitem[Merloni et al.(2003)]{Merloni03} Merloni, A., Heinz, S., 
\& di Matteo, T.\ 2003, \mnras, 345, 1057 

\bibitem[O'Dea(1998)]{Odea98} O'Dea, C.~P.\ 1998, \pasp, 110, 493 

\bibitem[Orienti \& Prieto(2010)]{Orienti10} Orienti, M., \&
Prieto, M.~A.\ 2010, \mnras, 401, 2599

\bibitem[Ott et al.(2005)]{Ott05} Ott, J., Walter, F., 
\& Brinks, E.\ 2005, \mnras, 358, 1423 

\bibitem[Pearson et al.(1992)]{Pearson92} Pearson, T.~J., 
Blundell, K.~M., Riley, J.~M., \& Warner, P.~J.\ 1992, \mnras, 259, 13P 

\bibitem[Peterson et al.(2005)]{Peterson05} Peterson, B.~M., et 
al.\ 2005, \apj, 632, 799 

\bibitem[Phillips et al.(2007)]{Phillips07} Phillips, C.~J., 
Deller, A., Amy, S.~W., et al.\ 2007, \mnras, 380, L11 

\bibitem[Reed et al.(1995)]{Reed95} Reed, J.~E., Hester, 
J.~J., Fabian, A.~C., \& Winkler, P.~F.\ 1995, \apj, 440, 706 

\bibitem[Reines et al.(2008a)]{Reines08a} Reines, A.~E., Johnson, 
K.~E., \& Goss, W.~M.\ 2008, \aj, 135, 2222 

\bibitem[Reines et al.(2008b)]{Reines08b} Reines, A.~E., Johnson, 
K.~E., \& Hunt, L.~K.\ 2008, \aj, 136, 1415 

\bibitem[Reines et al.(2010)]{Reines10} Reines, A.~E., Nidever, 
D.~L., Whelan, D.~G., \& Johnson, K.~E.\ 2010, \apj, 708, 26 

\bibitem[Reines et al.(2011)]{Reines11} Reines, A.~E., Sivakoff, 
G.~R., Johnson, K.~E., \& Brogan, C.~L.\ 2011, \nat, 470, 66 

\bibitem[Skrutskie et al.(2006)]{Skrutskie06} Skrutskie, M.~F., et 
al.\ 2006, \aj, 131, 1163 

\bibitem[Shepherd(1997)]{Shepherd97} Shepherd, M.~C.\ 1997, 
Astronomical Data Analysis Software and Systems VI, 125, 77

\bibitem[Thornton et al.(2008)]{Thornton08} Thornton, C.~E., 
Barth, A.~J., Ho, L.~C., Rutledge, R.~E., 
\& Greene, J.~E.\ 2008, \apj, 686, 892 

\bibitem[Ulvestad et al.(2007)]{Ulvestad07} Ulvestad, J.~S., 
Johnson, K.~E., \& Neff, S.~G.\ 2007, \aj, 133, 1868 

\bibitem[van Wassenhove et al.(2010)]{vanWassenhove10} van Wassenhove, 
S., Volonteri, M., Walker, M.~G., \& Gair, J.~R.\ 2010, \mnras, 408, 1139 

\bibitem[Volonteri et al.(2008)]{Volonteri08} Volonteri, M., 
Lodato, G., \& Natarajan, P.\ 2008, \mnras, 383, 1079 

\bibitem[Wrobel et al.(2001)]{Wrobel01} Wrobel, J.~M., Fassnacht, C.~D., \& Ho, L.~C.\ 2001, \apjl, 553, L23 

\end{thebibliography}
\end{document}